# Three-dimensional Kinematic Metamaterial

# with tuneable directional permeability

Yunfang Yang[1], Perla Maiolino[1,2], Yan Chen[3,4]* and Zhong You[1]*

1. Department of Engineering Science, University of Oxford, Parks Road, Oxford, OX1 3PJ, UK
2. Oxford Robotics Institute, University of Oxford, Parks Road, Oxford, OX1 3PJ, UK
3. Key Laboratory of Mechanism Theory and Equipment Design of Ministry of Education, Tianjin University, Tianjin 300072, China
4. School of Mechanical Engineering, Tianjin University, Tianjin University, Tianjin, 300072, China

* Corresponding authors. Email: yan_Chen@tju.edu.cn and zhong.you@eng.ox.ac.uk

**Abstract**

Reconfigurable metamaterials are constructed from tessellation of deformable modules that give rise to a set of tuneable properties. To date, most research focuses on metamaterials that morph between final configurations along a single deformation path. Multi-pathway metamaterials that can transform into several different shapes according to various external stimuli are rare. Here we propose a kinematic based metamaterial with multiple predefined deformation paths. The inherent kinematic bifurcations of the deformable unit cells enable the metamaterial to switch among different paths under simple stimuli at one particular configuration, resulting in the enhancement of reconfigurability. The metamaterials demonstrated in this paper can provide multiple open channels in different orthogonal directions, making them suitable for multifunctional applications such as tuneable filters and reconfigurable wave guiding materials.

**Main text**

Reconfigurable metamaterials are man-made structural materials with deformable internal constituents and topologies which can be tuned by embedded actuation or external stimuli(*1–5*). The large deformation of the metamaterials enables a range of properties that are tuneable, such as generating frequency shifts and loss changes in electromagnetic fields (*6*), and inducing anisotropy and chirality property for tuneable photonic application (*7, 8*). Compared to electrical and magnetic tunability, mechanical tuning provides a broad tuning range with lower electromagnetic loss and comparatively low fabrication cost (*6*), making them good candidates for tuneable filters and antennas (*9, 10*).

The capability of reconfiguration mainly arises from the geometry design of the constituent structural module of the metamaterials. In general the structural modules can be either purposefully engineered materials, e.g., materials with porous topology (*2, 11–14*), trusses whose elastic members buckle under external load and subsequently recover when unloaded (*15, 16*) and bi-stable units (*17*), or transformable cells with in-built motion capability, e.g., the kirigami (*18*) or origami (*19–21*) based modules and re-entrant cell (*22–24*). The latter, which is the main focus of this paper, are in fact mechanisms: assemblies of relatively rigid members that are able to undergo large shape morphing. For instance, some 2D metamaterials with negative Poisson's ratio are made of planar mechanisms in which rigid bodies are joined together by the flexible hinges enabling them to rotate with respect to each other while deforming(*25, 26*). The origami/kirigami modules are essentially a set of relatively rigid panels connected by rotational creases, which are mechanically equivalent to spherical or spatial mechanisms (*27–30*).

The advantage of mechanism-based metamaterials is their ability of undergoing reliable large deformation with distinct motion path, a feature that is critical to achieve robust tuneability and programmability. Their performance is, in general, independent of the base materials used, making much wider applications of such concepts possible (*31, 32*). To date, most mechanism-based metamaterials have only one motion path. For example, auxetic structures often morph between two final configurations (*7, 26*) in spite that multi-pathway metamaterials that can morph into multiple shapes according to different external stimuli are more suitable for multifunctional applications. Efforts have been made to increase the number of motion paths using either multi-step or multi-mode designs. The former (*33*) provides a sequence of configurations when the material is subjected to uniform compression, yet the order of these configurations is pre-determined. The latter (*34*) is metamaterial made from interlinked extruded cubes that gives rise to multiple degrees of freedom. Each structural module has to be manipulated by several actuators in order to reach a particular configuration, making the control of deformation rather complex.

Therefore, in this paper, we propose a kinematic bifurcation based reconfigurable metamaterial with multiple predefined and reliable deformation paths. Kinematic bifurcation is a point of intersection of different motion paths. The mechanism can only switch its motion path at this particular configuration, and once it left this point the motion is fully determined by the chosen path. Previous research on deployable structures mainly try to avoid any kinematic bifurcation (*35*), while here we exploit it to achieve material reconfiguration for the first time. In our approach, spatial kinematic mechanisms with exact deformation paths are entrenched within the construction cell of the metamaterial, and the cell deforms into various desirable shapes through their inherent kinematic bifurcations. These unit cells are subsequently tessellated in three orthogonal directions to create truly 3D metamaterials, whose topology is guided by the kinematics so that the tuneability of the cells is strictly maintained. Our strategy ensures that such metamaterials can transform precisely into different shapes under simple stimuli due to mechanical coupling. Examples in this paper demonstrate that we can create 3D metamaterials that provide single or multiple open channels in one of more of the orthogonal directions, which can be used to regulate air or fluid flows or to direct sound waves.

**Results**

We begin by connecting six prisms edge by edge to form a closed loop, as shown in Fig. 1a. The prisms are connected in a way where two sets of parallel connection edges facing different directions. The four blue prisms in the middle are in cube shapes, and the red top and bottom rectangular prisms are twice the size of the blue cubes to avoid physical collision during motion. For convenience, let us refer the rectangular prisms as end blocks, and the middle cubes as tie blocks. This assembly forms a Sarrus linkage capable of rectilinear motion where two end blocks can move vertically up and down relative to each other whereas the left and right pairs of tie blocks swing closer or apart. This six-block assembly is taken as our basic kinematic element.

Note that the tie blocks are connected on two edges in diagonal corners of the cube, therefore, the equivalent bars of the Sarrus linkage are the yellow panels shown in Fig. 1a. In general, the Sarrus linkage has one degree of freedom. However, when the angles between the two adjacent tie blocks, $\theta_x$ and $\theta_y$, become $\pi/2$, i.e., the two yellow panels in the same set are coplanar with each other, the two tie blocks can make a change to their swing direction. In another word, it is possible for $\theta_x < \pi/2$ whilst $\theta_y > \pi/2$, or $\theta_x < \pi/2$ and $\theta_y < \pi/2$, which leads to two different deformation paths (see S1 in Supplement text). This particular configuration is a kinematic bifurcation point(*36*). Kinematic bifurcation is different from multi-DOF assemblies because the former can only switch its motion paths at the bifurcation point. After a specific path is chosen, the motion can again be fully described by a single input. The selection of different motion path can be guided by a perturbation at the bifurcation point using external stimuli.

A kinematic unit is obtained by connecting four such elements through intermedium planer 4*R* linkages, which join the parallel edges in the pair of tie blocks with the corresponding edges in the adjacent element, as shown in Fig. 1b. The unit has a single DOF because the rectilinear motions of all elements are synchronized by the intermedium linkages. If we choose $\theta_x$ as the kinematic variable, we can derive different angle relations indicating four independent motion paths:

$$P1: \theta_x = \theta_y, \theta_x \in [0, \pi/2],$$

$$P2: \theta_x = \pi - \theta_y, \theta_x \in [0, \pi/2],$$

$$P3: \theta_x = \theta_y, \theta_x \in [\pi/2, \pi],$$

$$P4: \theta_x = \pi - \theta_y, \theta_x \in [\pi/2, \pi]. \tag{1}$$

Noting that the motion range for $\theta_x$ and $\theta_y$ is $[0, \pi]$, and the bifurcation point is at $\theta_x = \theta_y = \pi/2$ where two paths cross. Other angles can all be derived by $\theta_x$ (see S1 in Supplementary text). The kinematic deformation paths are displayed in Fig. 1c together with their corresponding shapes.

Looking from any of the *x*, *y* and *z* directions, the outline of the unit always has a rectangular shape. Therefore, we can tessellate the unit in a grid pattern in three directions to construct a 3D kinematic metamaterial shown in Fig. 1d.

Fig. 1e shows a kinematic metamaterial prototype consisting of 2×2×2 units fabricated using multi-material additive manufacturing with flexible material for connection hinges (1 mm thickness) and stiffer material for cubes (edge length 12 mm). Its motion was mechanically tested under uniaxial compression in $z$-direction using a transparent loading plate. The experimental setup is shown in the Materials and Methods. Performance of the prototypes of this metamaterial is illustrated in Movie S2.

If we consider that the unit has four deformation paths intersecting at one bifurcation point in Fig. 1b, the 2×2×2 bifurcated metamaterial has nine deformation paths which are controlled by $\theta$ angles in each unit. To show these paths more clearly, we plot them with respect to the engineering strains in two orthogonal directions $x$ and $y$ in Fig. 2a. During the deformation, it is observed that in the same row along the $x$ axis of one layer, the width alterations of each unit are always the same. Likewise, the width of each unit of the same column along $y$ direction is always equal. Therefore, starting from the bifurcation point X(Fig. 2b), the metamaterial can, for instance, become fully packed without any voids (I), expand only four units of same column in $x$ direction forming two open channels (II), or expand all eight units in both columns in $x$-direction while keeping $y$-direction firmly shut (III). In total, the metamaterial can realise nine distinct final states, I – IX, which are displayed in Fig. 2a. To transfer from one final shape to the other, it is necessary to pass through the singular state X of bifurcation.

The final states have the following two distinctive features. First, open channels can be created column by column, which is evident by II and III with channels in $x$ direction only, and IV and VII where channels appear in $y$ and $z$ direction. This feature can be extended to states where orthogonal channels appear such as in V, VIII, VI and IX. Secondly, II and III provide open channels in $x$ direction while no voids appear in either $y$ or $z$ directions, which indicates that creation of open channels in $x$ directions is completely decoupled with those in $y$ and $z$ directions. However, the appearance of channels in $y$ direction, as shown in IV and VII, is accompanied by channels in $z$ direction, meaning that $y$ and $z$ are coupled. In the states V, VI, VII, IX, there are open channels in all three orthogonal directions. The creation of open channels indicates that this kinematic metamaterial offers multi-directional and controllable permeability through varying the number of opened channels and their directions. The elastic energy to trigger each deformation path is the same, so there is no preference for path selection at the bifurcation point.

The engineering strains of the 3D printed model match well with our theoretical predictions, as shown in Fig. 2e. The porosity $p$ of the metamaterial is plotted in Fig. 2f. To quantify the permeability of the material in three orthogonal directions, we also calculate its void fraction $f_x$, $f_y$ and $f_z$, the ratio between the cross-sectional areas of the void and the entire metamaterial in $x$, $y$ and $z$ directions, respectively. The results are given in Fig. 2g (see S2 in Supplementary text). As we discussed before, the void fraction at nine terminal states I and IX are most interesting. For example, at state II, $f_x = 0.2$ whereas $f_y = f_z = 0$, indicating that there are only channels in $x$-direction. At state IV, $f_x = 0$, and $f_y = f_z = 0.18$, which means channels in $y$ and $z$ directions co-exist.

To guide the metamaterial to a desired path, we design several pairs of control plates and place them on the top and bottom of the metamaterial. As shown in Fig. 3a, the transparent control plates have different columns and rows of bump to block certain channels of the metamaterial, leading to deformations towards different motion paths. This is possible because during the deformation, the end blocks will be led to move towards one direction among northwest, northeast, southwest and southeast. If we manually block three directions, it will only move towards the rest one. For instance, to reach state V, we design a pair of control plates with grid profile so that one channel in both $x$ and $y$ directions are blocked. As shown in Fig. 3b to 3d, the end prisms in the blocked unit can no longer move inward, they can only translate outward until reach the final state V. On the other hand, we use plates with hollow rectangular profile to reach state I. The metamaterial cannot expand itself in either *x* or *y* directions, so each unit will shrink along Path 2 until the metamaterial reaches state I. In Fig. 3e, by designing the position of bumps we block three paths of each unit everytime, leaving only one passible deformation direction, and the selection of deformation paths is therefore settled. The strength of this control method is that there are no actuators on the metamaterial to trigger the switch of deformation path. The control plates are mounted on the operation table in such a way that they can hold the material in between, then the deformation is actuated by a single compression input on the control plate. Therefore, it can function as a nonelectronic tunable filter in fluid or other complex environments to guide the flow. For applications that require no cover plates on the top and bottom of the metamaterial, the switch can also be triggered by embedded actuators made from shape memory polymers(*37*), heat driven(*38*) or pneumatic hinges(*39*).

Should the number of units be increased, the number of deformation paths and distinct states will increase, but the fundamental behaviour remains unchanged. For instance, for a metamaterial with $k_x \times k_y \times k_z$ units (in *x*-, *y*- and *z*-directions), there will be $(k_x+1) \times (k_y+1)$ distinct final states.

**Discussion and conclusions**

A kinematic bifurcation based reconfigurable metamaterial is presented here for the design of 3D multi-path metamaterials. The unit cell of the resultant metamaterial is constructed from a single degree of freedom spatial mechanism with a kinematic bifurcation point, and it subsequently tessellated to obtain the structural backbone of the metamaterials while preserving the geometrical compatibility among the mechanisms. The metamaterial has multiple deformation paths inherited from their unit cells, and the overall single DOF of their frameworks ensures that the deformation of metamaterial can follow the predefined kinematic motion path without external synchronisation. Because of the existence of kinematic bifurcation, the resultant metamaterial exhibits the capability to tune certain property of the metamaterials by switching amongst different prescribed motion paths. The only steering required is at the bifurcated state, which can be achieved many stimuli such as forces applied externally or thermal/magnetic-induced actuation. In our experiments, we simply used physical obstructions to block other motion paths at the bifurcation point.

The bifurcation based reconfigurable metamaterials are scale and base material independent, and they utilise only flexible rotational connections amongst construction blocks, making it possible to fabricate or synthesise them in a wide range of scales and base materials. The stiffness of our materials through the deformation phase is provided by the stiffness of rotational connections. All of our prototypes had highly reliable performance with very low sensitivity to both base material and manufacture errors.

The kinematic unit in this paper is based on the Sarrus linkage. We envisage that the proposed design strategy for kinematic metamaterial can be extended to units using other 3D mechanisms. The metamaterials with entrenched accurate deformation behaviour that can be turned on/off, or changed by external stimuli can have many applications which we are actively exploring.

This research paves the way for a new class of multi-path 3D kinematic metamaterials that can perform complex and reliable reconfiguration, bringing metamaterials closer to multi-tasks applications arising in areas such as stealth technology(*40*), reconfigurable acoustic(*41*, *42*) and photonic(*4*) metamaterials, micro-scale tuneable filters(*5*, *43*), soft robots(*44*, *45*) and wearable technology(*3*).

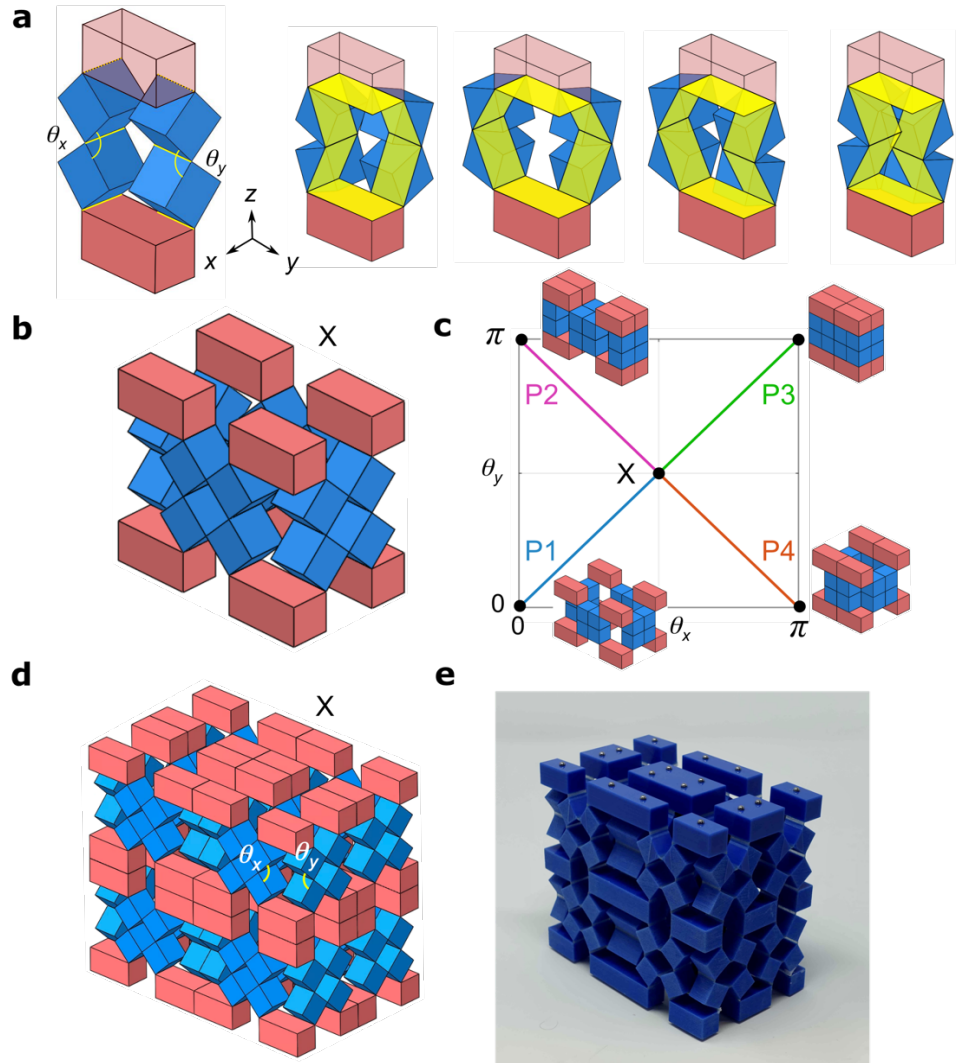

**Figure 1.** Construction of a reconfigurable metamaterial with kinematic bifurcation. (a) The 1 DOF six prism element based on Sarrus linkage. (b) A unit cell constructed by four elements. The angular relationship indicates that the unit has four motion paths intersect at the bifurcation point. (c) A 2×2×2 unit metamaterial obtained by tessellating the unit in three orthogonal directions and the 3D-printed model.

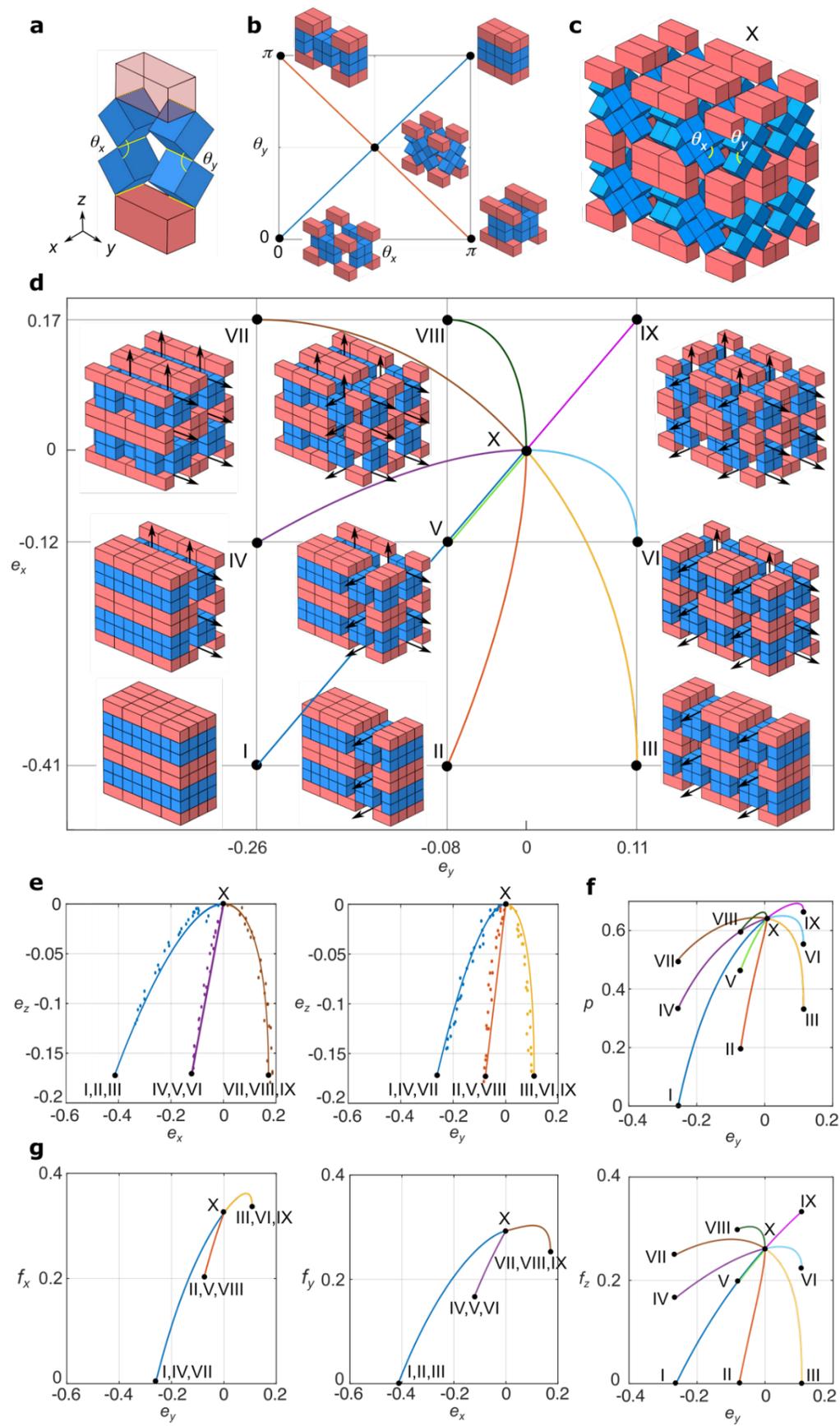

**Figure 2.** The kinematic metamaterial with tuneable multi-directional permeability. (a) The element with double-sized end blocks. (b) The unit created by four elements has a kinematic bifurcation point at which a switch is permitted in its deformation paths. (c) A 2×2×2 unit bifurcated kinematic metamaterial at the bifurcation point, state X. (d) Nine deformation paths of the metamaterial in $e_x$ vs. $e_y$ plot with the corresponding final states at the end of each path. The metamaterial can deform into any of the nine states (I to IX) through state X. The channel directions in each state are shown by black arrows. (e) Theoretical engineering strain plots (solid curves) and experimental validation (dots). (f) Porosity $p$ vs. $e_y$. (g) Void fractions $f_x$ vs. $e_y$, $f_y$ vs. $e_x$ and $f_z$ vs. $e_y$.

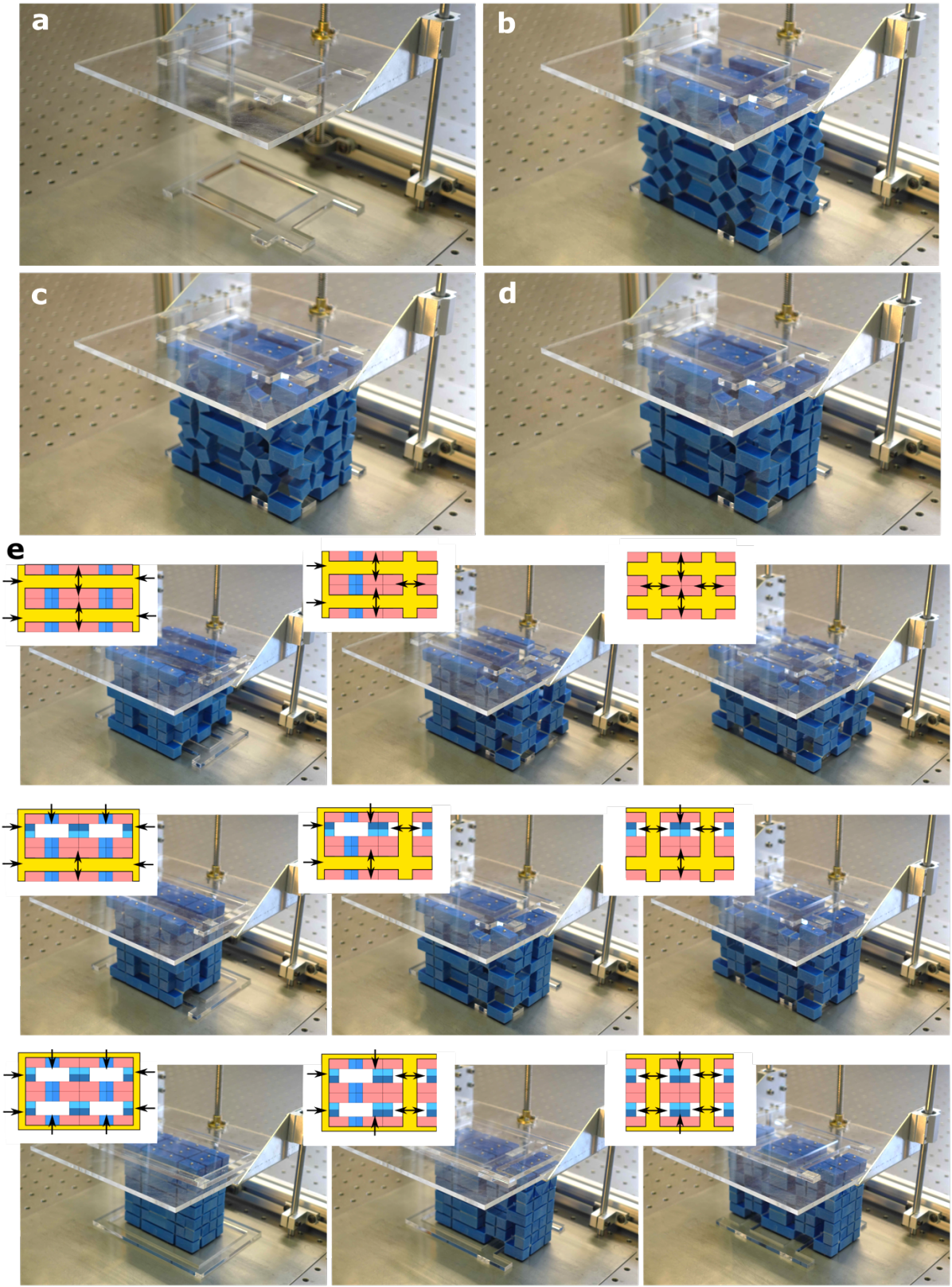

**Figure 3.** Control method of the bifurcated kinematic metamaterial. (a) The control plates for state V are mounted on the operation table. (b-d) The deformation process of the kinematic metamaterial under the guidance of control plate. (e) Nine control plates designed for different paths of the metamaterial with their corresponding final states at the end of each path. Initially, the metamaterial is in state X. It is placed in between the control plates, and the deformation is triggered by compression force on the plates.

# Supplementary Materials for

# Three-dimensional Kinematic Metamaterial with tuneable directional permeability

Yunfang Yang, Perla Maiolino, Yan Chen* and Zhong You*

*Correspondence to: zhong.you@eng.ox.ac.uk and yan_Chen@tju.edu.cn

**This PDF file includes:**

- Materials and Methods
- Supplementary Text
  - S1 Kinematic analysis of spatial linkages
  - S2 Physical properties of the kinematic metamaterials
- Figs. S1 to S5
- Captions for Movies S1 and S2

**Other Supplementary Materials for this manuscript include the following:**

Movies S1 and S2

# Materials and Methods

## Fabrication of the prototypes

Two manufacturing methods were developed for producing prototypes of the kinematic metamaterials. The first was a multi-process technique which involves assembling parts produced by 3D printing and laser cutting. The steps taken are demonstrated in Fig. S1a. We started by printing each face of the cube building block with snap-fit clips (two male panels and four female panels), and by manufacturing connections using laser-cut 0.05mm mylar sheet with holes corresponding to those extruded holders on the clip. Then the mylar pieces were attached to the extruded holders on face panels at the edges intended for connection with neighbouring cubes. Finally, using the snap-fit clips, we mechanically held the cubes together, one after another, to assemble the unit prototype. The mylar pieces clamped by two adjacent cubes acted as connection hinges. The cube building blocks are rigid, and they were able to rotate freely with respect to each other for the mylar connections offered little resistance, making it perfect to assess the kinematic deformation of the prototype.

The second method utilised an additive manufacturing technique to print the entire metamaterial prototype in a single step, which was done by a multi-material 3D printer (Stratasys J735). Relatively rigid material (Verocyan RGD841, Young's modulus 2000–3000MPa, Tensile strength 50–65MPa) was used for cubes (8mm edge length) and flexible but durable material (Agilus30 Clear FLX935, Tensile strength 2.4–3.1MPa) was adopted for the connections. A parametric study of the hinges was performed that resulted in the optimum dimensions of the connections being $w = 1$mm and $l = 1.5$mm, as shown in Fig. S1b. The metamaterial was made in its bifurcation state when $\theta_x = \pi/2$.

## Experiments

To measure the strain relations of the metamaterial, tests were performed by placing the sample between a rigid metal plate and an acrylic plate, see Fig. S2. The latter was allowed to move up and down along the guiding rails once pressed. For the assembled prototype, there was little friction between its top and bottom surfaces and the two plates, so the prototype shrank or expanded freely under compression. For multi-material printed samples, however, the friction between the prototype and plates could be considerable, which would adversely affect the deformation behaviour of the model. To reduce friction, we embedded a number of 2mm diameter steel balls on the top and bottom surfaces of the end blocks, as shown in Fig. S2a.

The 3D printed prototype was placed at its initial printed configuration to start with. The acrylic board was then slowly lowered by compression in z-direction, and the deformation of the metamaterial was recorded by a video camera in $x$-, $y$-, $z$-directions. We tested each motion path separately for three times and obtained a combined plot in Fig. 2e. Then the videos were processed as a series of snapshots, and the length variations in $z$-direction and $x$- or $y$-directions were measured by post-processing the images. Finally, the engineering strains were calculated, which were then used to obtain engineering Poisson's ratios.

**Supplementary Text**

**S1 Kinematic analysis of spatial linkages**

**S1.1** *The matrix method using DH notation*

The number of degrees of freedom (DOFs), also known as the mobility of a mechanism, is the number of input motion required to determine the position of all the rigid bodies with respect to the ground. For an assembly consisting of *n* rigid bodies (links), *j* joints, and the DOFs of each joint $f_i$ ($i$ =1, 2,…, *j*), the mobility *m* can be calculated by the Grubler-Kutzbach mobility criterion(*56*):

$$m = 6(n-j-1) + \sum_{i=1}^{j} f_i \qquad (S1)$$

Here we focus on linkages formed by *n* links connected by *n* hinges (also known as revolute joints where $f_i$ = 1). Using Eq. (S1), it can be found that $m = n - 6$. In general, *n* must be greater than 6 to enable mobility. If *n* is less or equal to 6, the assembly is either a rigid structure or an overconstrained mechanism, a transformable mechanism where *m*, calculated using Eq. (S1), is less than one(*57*). The overconstrained mechanism exists usually under special geometrical conditions. Its motion can be analysed using the matrix method with the Denavit and Hartenberg notation (DH notation)(*58*). When a set of *n* coordinate systems being set up, one for each of the joints, under the DH notation as shown in Fig. S3, the closure equation of such linkage is

$$\mathbf{T}_{12}\mathbf{T}_{23}\cdots\mathbf{T}_{n1} = \mathbf{I}_4 \qquad (S2)$$

where $\mathbf{I}_4$ is the 4×4 unit matrix, and $\mathbf{T}_{i(i+1)}$ ($i$ = 1, 2, …, *n*. If $i+1 > n$, it is replaced by 1) is the transformation matrix with the following form.

$$\mathbf{T}_{i(i+1)} = \begin{bmatrix} \cos\theta_i & \sin\theta_i & 0 & -a_{i(i+1)} \\ -\cos\alpha_{i(i+1)}\sin\theta_i & \cos\alpha_{i(i+1)}\cos\theta_i & \sin\alpha_{i(i+1)} & -R_i\sin\alpha_{i(i+1)} \\ \sin\alpha_{i(i+1)}\sin\theta_i & -\sin\alpha_{i(i+1)}\cos\theta_i & \cos\alpha_{i(i+1)} & -R_i\cos\alpha_{i(i+1)} \\ 0 & 0 & 0 & 1 \end{bmatrix} \qquad (S3)$$

In this equation, $a_{i(i+1)}$, length of the link, $\alpha_{i(i+1)}$, the twist angle between joints *i* and *i* +1, and $R_i$, the offset of joint *i*, are the geometrical parameters, whereas $\theta_i$, the rotational angle between the two links connected at joint *i*, is the kinematic variable, all of which are shown in Fig. S3.

For a given set of geometrical parameters, if Eq. (S2) holds for any kinematic variable $\theta_i$ within certain range, the linkage has at least one DOF.

## S1.2 Kinematic analysis for the Metamaterial

The mechanism entrenched in the element of the metamaterial is a special case of the Sarrus linkage, which is an overconstrained $6R$ spatial mechanism consisting of a closed chain of six rigid bodies connected by six revolute joints(*59*). The mechanism, reproduced in Fig. S4(a), has

a set of geometrical parameters that satisfy the following conditions:

$$\alpha_{12} = \alpha_{23} = \alpha_{45} = \alpha_{56} = 0, \ \alpha_{34} = \alpha_{61} \ (\alpha_{34} \neq 0, \pi, 2\pi),$$
$$a_{12} = a_{23} = a_{45} = a_{56} = \sqrt{2}a, \ a_{34} = a_{61} = 0. \tag{S4}$$

as shown in Fig. S4(b). Two sets of solutions can be obtained using Eq. (S2), which are

$$\theta_1 = -\frac{\theta_5}{2}, \ \theta_2 = \theta_5, \ \theta_3 = \theta_1 - \pi, \ \theta_4 = \theta_1, \ \theta_6 = \theta_3, \tag{S5}$$

and

$$\theta_1 = \frac{\theta_5}{2}, \ \theta_2 = 2\pi - \theta_5, \ \theta_3 = \pi - \theta_1, \ \theta_4 = \theta_1, \ \theta_6 = \pi - \theta_1. \tag{S6}$$

If we use the dihedral angles $\theta_x$ and $\theta_y$, there are $\theta_x = \theta_5 - \frac{3\pi}{2}$ and $\theta_y = \theta_2 - \frac{3\pi}{2}$. Equations (S5) and (S6) then give

$$\theta_x = \theta_y \text{ and } \theta_x = \pi - \theta_y, \tag{S8}$$

respectively. Both are plotted in Fig. 2b. Noting that the motion range for $\theta_x$ and $\theta_y$ is [0, π], and there is a bifurcation point at $\theta_x = \theta_y = \frac{\pi}{2}$ where two paths cross.

## S2 Physical properties of the kinematic metamaterials

### S2.1 Directional permeability of the metamaterial

We have explained that the bifurcated metamaterial is able to generate channels in some directions to enable liquid or gas to pass while close the voids in other directions to prevent the flow. To quantify the permeability of the material from different directions, we calculate its void fraction $f_{ij}$, the ratio of the cross-sectional area of the material which is occupied by openings,

$$f_{ij} = 1 - \frac{A_{s-ij}}{A_{t-ij}} \ (i, j = x, y, z \text{ and } i \neq j)., \tag{S29}$$

in which $A_{s-ij}$ denotes to the surface area of the solid building blocks, and $A_{t-ij}$ the total surface area of the metamaterial in the $xy$, $zx$ or $yz$ planes. This term is mainly used in gas-liquid two-phase flow, and here we adopt it to demonstrate the permeability in three orthogonal directions.

For an $n \times n \times n$ unit metamaterial, along each path, denote $q$ and $p$ as the numbers of columns in $x$-direction and $y$-directions that contain open channels, respectively. In $x$-direction, let $\theta_y$ be the dihedral angle in $q$ columns, and the rest of the columns that are closed will have dihedral angle $\pi - \theta_y$. Similarly, in $y$- direction, let $\theta_x$ be the dihedral angle in $p$ columns, the rest of the columns that are closed will have dihedral angle $\pi - \theta_x$. Note that $\theta_x = \theta_y$. The surface area of the solid building blocks can then be calculated as

$$A_{s-xz(p,q)} = 2naL_z + 2(n-q)\tan\frac{\theta_y}{2}a^2 + 2q(2 - \tan\frac{\theta_y}{2})a^2,$$

$$A_{s-yz} = 2naL_z + 8na^2, \tag{S30}$$

$$A_{s-xy} = 2naL_x + 2naL_y - 4n^2a^2.$$

The total surface area is simply

$$A_{t-ij} = L_i \cdot L_j \ (i, j = x, y, z) \tag{S31}$$

where the dimensions of the metamaterial $L_x$, $L_y$, $L_z$ are given by

$$L_y = 4na + p \cdot 2a \sin\frac{\theta_x}{2} + (n-p) \cdot 2a \sin(\frac{\pi - \theta_x}{2}),$$

$$L_x = 2na + q \cdot 2a \sin\frac{\theta_y}{2} + (n-q) \cdot 2a \sin(\frac{\pi - \theta_y}{2}), \tag{S32}$$

$$L_z = 2na + 2na(\sin\frac{\theta_x}{2} + \cos\frac{\theta_x}{2})$$

Once the surface areas are found, substituting them into Eq. (S29) gives the void fraction in three planes.

### S2.2 Porosity of the metamaterial

The porosity of the metamaterial during transformation, $p$, is

$$p = \frac{V_t - V_s}{V_t} \tag{S33}$$

where $V_s$ is the total volume of the solid cube blocks, which is $V_s = 32n^3a^3$. $V_t$ is the total volume of the material, which can be calculated by

$$V_t = L_x \cdot L_y \cdot L_z \tag{S34}$$

Substituting the volume $V_s$ and $V_t$ into Eq. (S33) gives the porosity of the material.

### S2.3 Coupling of directional permeability of the metamaterial

In the kinematic metamaterial in Fig. 2, the formation of directional channels in $x$- direction is decoupled with those in $y$- and $z$- directions in any of the nine final states, while those in $y$- and $z$- directions are coupled. The latter indicates that, if the metamaterial deforms into a state with open or closed channels in $y$- directions, the channels will also open or close in $z$- direction, as shown in Figure S5a.

The coupling or decoupling of formation of directional channels can be tuned by changing the shape of end blocks. For instance, if we adopt the end blocks twice of the size of the current one, as shown in Fig. S5b, the formation of channels in $z$- direction is coupled with either $x$- or $y$- directions whilst those in $x$- and $y$-directions are decoupled. This indicates that the creation of channels in either $x$- or $y$- directions will also lead to channels in $z$- direction. However, if the elements are arranged in a unit by rotational symmetry, Fig. S5c, there are always channels in $z$-direction, which is decoupled with the formation of channels in either $x$- or $y$-directions. Meanwhile, the formation of channels in $x$- direction is decoupled with that in $y$- direction, and vice versa.

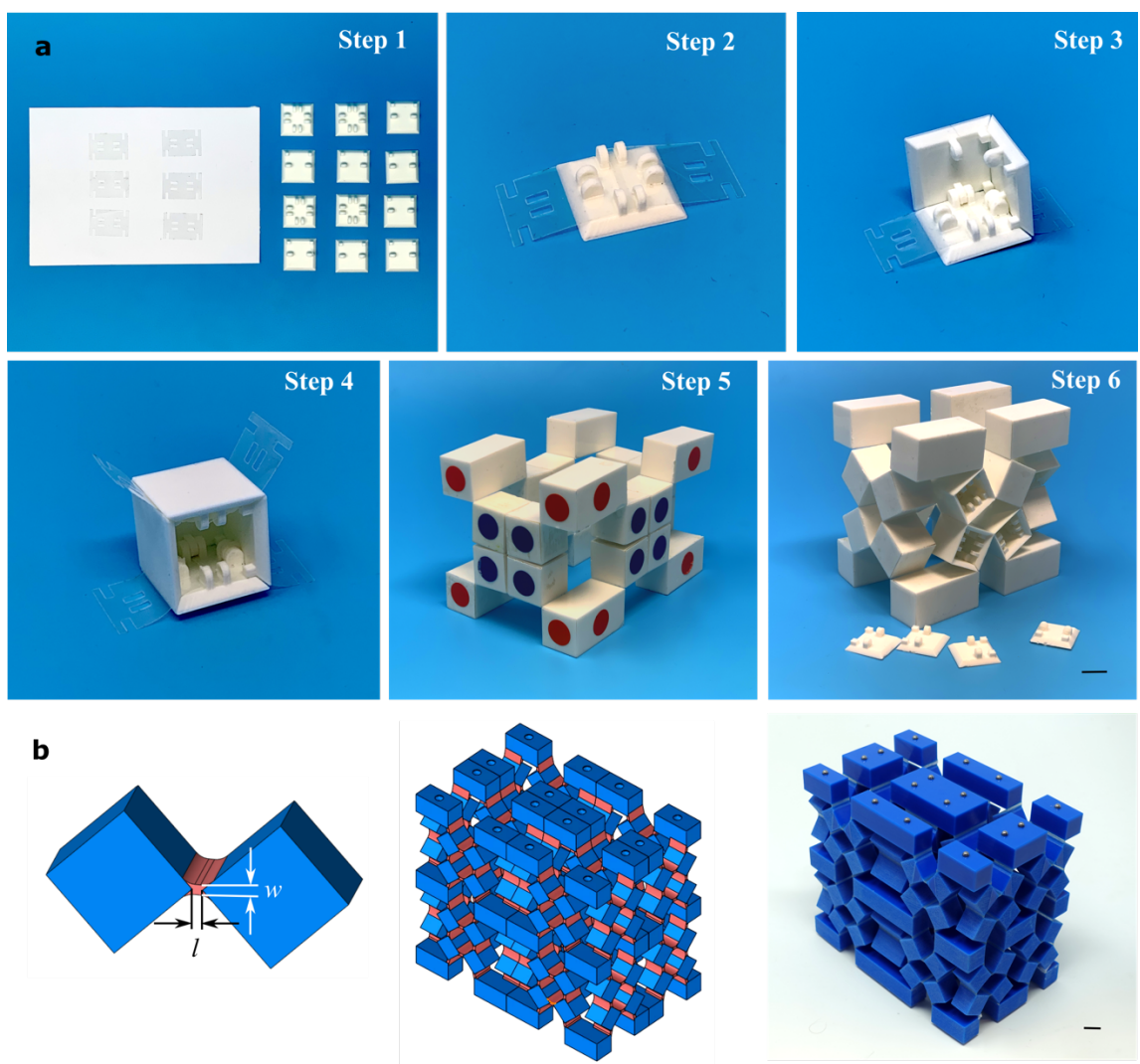

**Fig. S1.**

Fabrication of the prototypes. (a)A multi-process technique involves 3D printing and laser cutting of pieces (step 1), which are then assembled into interlinked cubes for the kinematic unit prototype (steps 2–6). (b) A multi-material printing technique. Scale bar size: 1cm.

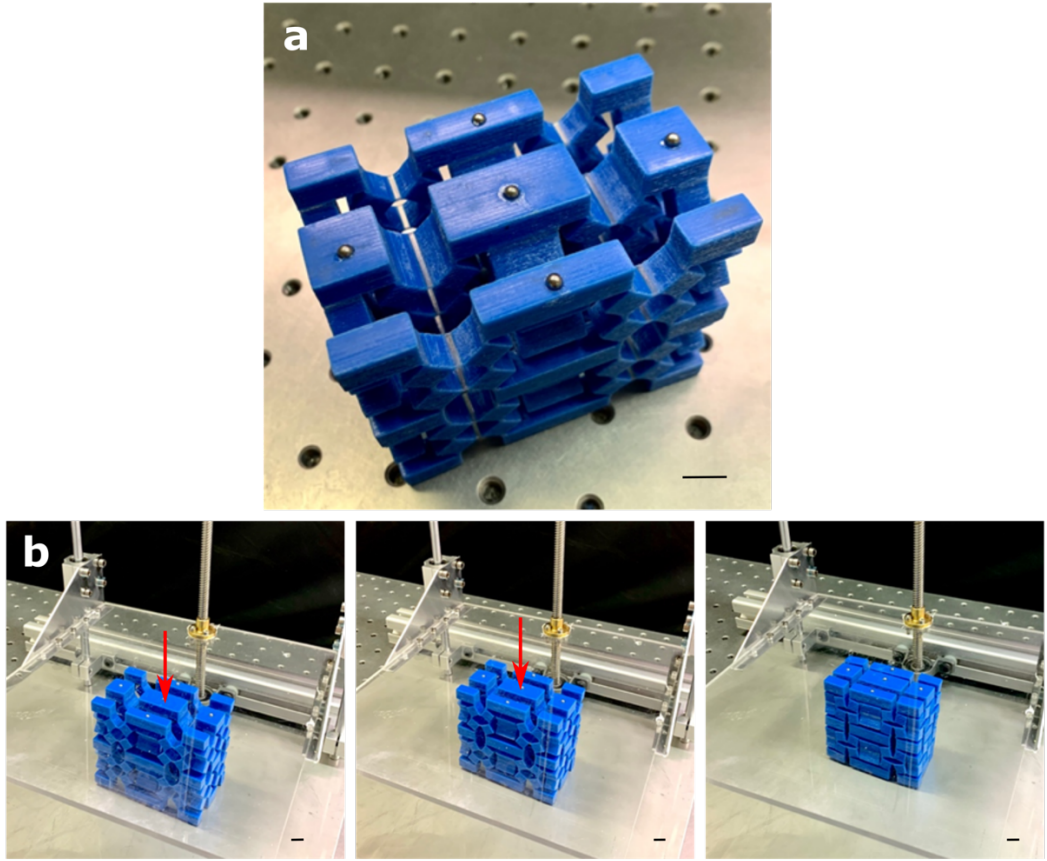

**Fig. S2.**

Experimental setup. (a) Stainless steel balls were embedded on the top and bottom surfaces of the metamaterial to reduce friction. (b) Compression progress of the metamaterial with multi-directional permeability. Scale bar size: 1cm.

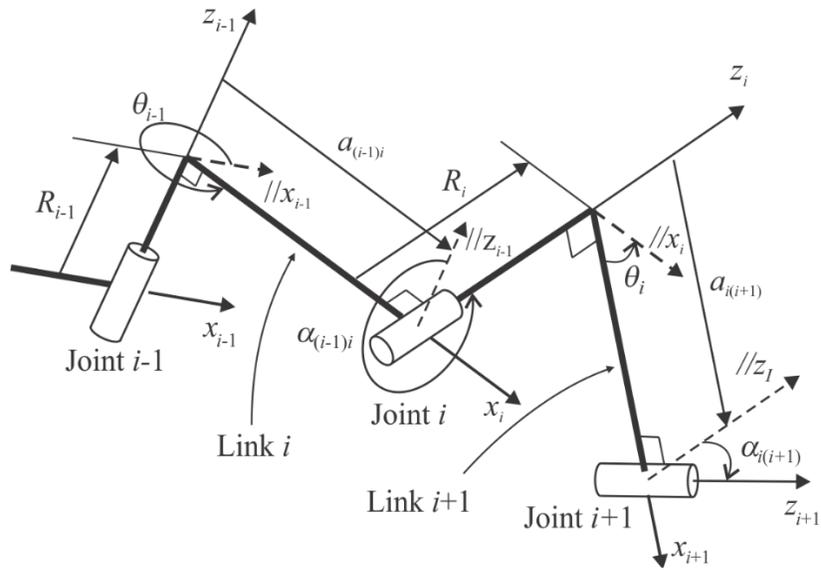

**Fig. S3.**
A portion of a linkage in which two links are connected by joint $i$. The coordinate systems are set up using DH notation where $z_i$ is placed along the axis of the joint $i$ and $x_i$ is along the length of link from joint $i-1$ to $i$.

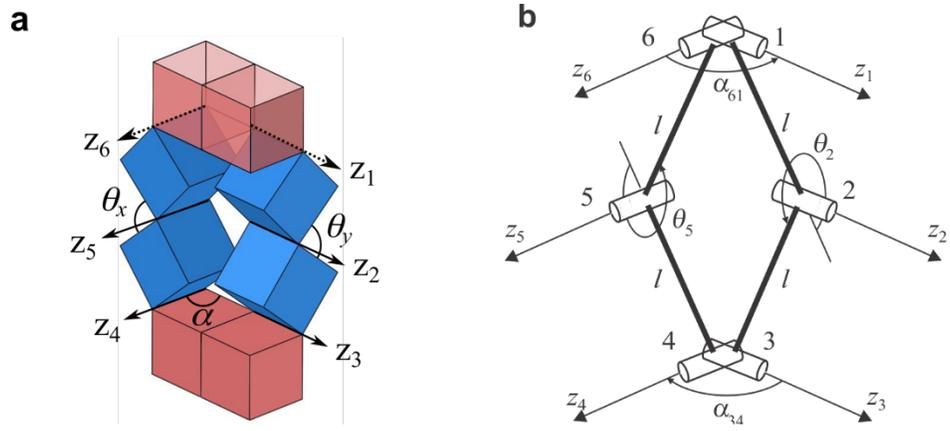

**Fig. S4.**
Kinematic analysis of the Sarrus linkage-based kinematic element. (a) Coordinate systems under DH notation to describe the motion of the element; (b) its equivalent Sarrus linkage in DH-notation;

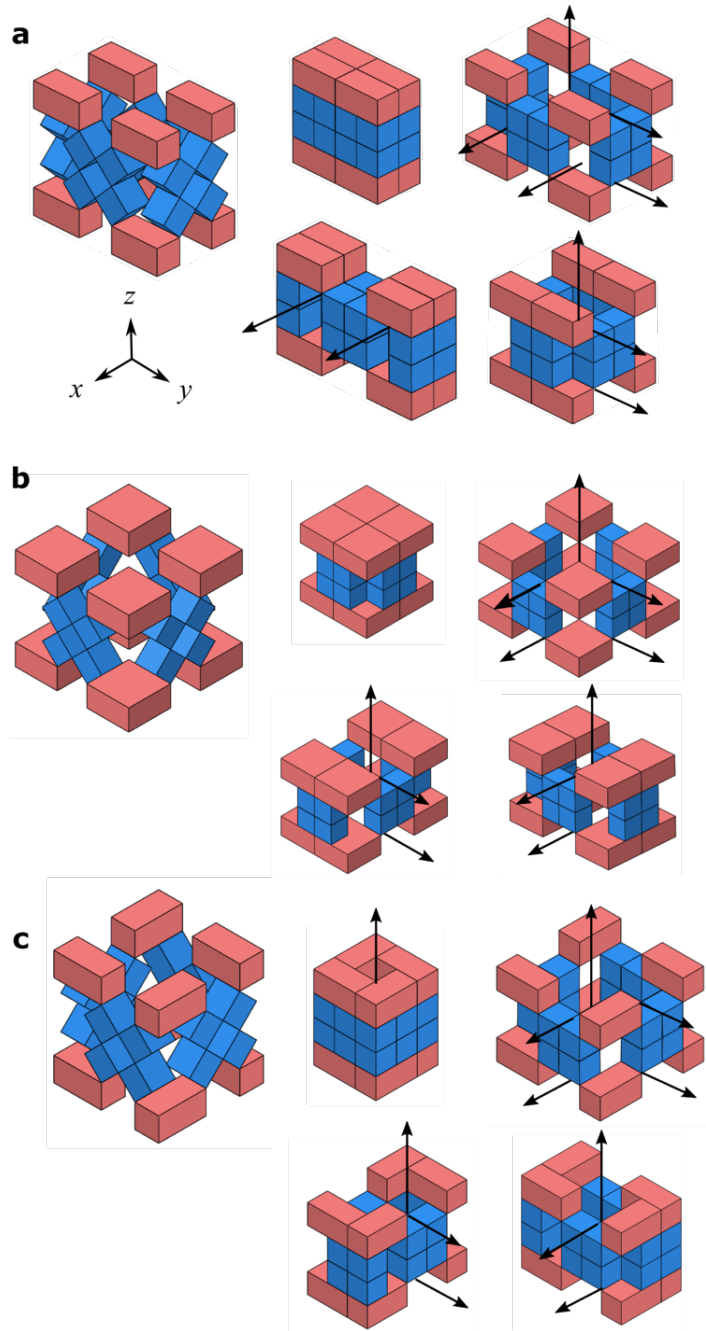

**Fig. S5.**
Coupling and decoupling of the formation of directional channels. (a) Coupling of channels in $y$- and $z$-directions. (b) Channel in $z$-direction is always coupled with that in either $x$- or $y$- directions. (c) Decoupling of channels in $x$-, $y$- and $z$-directions, but channel always exists in $z$-direction.

**Movie S1.**

Construction and Behaviour of the Metamaterial

**Movie S2.**

Bifurcation switch of the Kinematic Metamaterial